\documentclass[fleqn,usenatbib]{mnras}
\usepackage{newtxtext,newtxmath,xcolor, bm, subfig}
\usepackage[export]{adjustbox}

\usepackage[T1]{fontenc}
\usepackage{ae,aecompl}
\usepackage{graphicx}	
\usepackage{amsmath}	
\usepackage{amssymb}	
\title[Detecting SSOs with CNNs]{Detecting solar system objects with convolutional neural networks}
\author[M. Lieu et al.]{
Maggie Lieu,$^{1}$\thanks{E-mail: maggie.lieu@esa.int}
Luca Conversi$^{1}$
Bruno Altieri$^{1}$ and 
Beno\^it Carry$^{2}$
\\
$^{1}$European Space Astronomy Centre, ESA, Villanueva de la Ca$\tilde{n}$ada, E-28691 Madrid, Spain \\
$^{2}$Universit\'e C{\^o}te d'Azur, Observatoire de la C{\^o}te d'Azur, 
CNRS, Laboratoire Lagrange, France}
\date{Accepted XXX. Received YYY; in original form ZZZ}

\pubyear{2018}

\begin{document}

\label{firstpage}
\pagerange{\pageref{firstpage}--\pageref{lastpage}}
\maketitle

\begin{abstract}
In the preparation for ESA's Euclid mission and the large amount of data it will produce, we train deep convolutional neural networks on Euclid simulations to classify solar system objects from other astronomical sources. Using transfer learning we are able to achieve a good performance despite our tiny dataset with as few as 7512 images. Our best model correctly identifies objects with a top accuracy of 94$\%$ and improves to 96$\%$ when Euclid's dither information is included. The neural network misses $\sim$50$\%$ of the slowest moving asteroids (v $<$ 10 arcsec h$^{-1}$) but is otherwise able to correctly classify asteroids even down to 26 mag. We show that the same model also performs well at classifying stars, galaxies and cosmic rays, and could potentially be applied to distinguish all types of objects in the Euclid data and other large optical surveys.
\end{abstract}



\section{Introduction}
The solar system small bodies (asteroids, comets, Kuiper-belt objects 
(KBO)) are the remnants of the rocky and icy bodies that accreted to 
form the planets in the early solar system. Their orbital, size, and 
compositional distribution are the results of the mass removal and 
radial mixing triggered y the planetary migration in the early history 
of the solar system, and of Gyrs of collisions \citep{Bottke2002, Michel2015}

While their dynamics have provided the main constraints on the 
development of theoretical models over the last decade \citep[e.g.,][]{Morbidelli2005,Raymond2017} we 
are entering an era in which the compositional distribution of solar 
systems small bodies is maybe becoming even more important \citep{DeMeo2013, DeMeo2014}.

In particular, the populations of small to medium-sized KBO (tracers of 
the conditions in the outer planetary nebula) and small main belt 
asteroids (belonging to collisional families and hence progenitors of 
the near-Earth asteroids and meteorites) are too faint for current 
facilities \citep[e.g.,][]{LSST2009, Spoto2015}.

Whilst future large sky surveys such as LSST \citep{LSST2009} will likely uncover and characterise a large proportion of these undiscovered solar system bodies, ground based telescopes are limited to night-time observations and good seeing conditions. With an estimated launch in 2022, ESA's upcoming visible and near infrared space telescope Euclid \citep{Laureijs2011} is unlike many of the current surveys that typically focus on objects within the ecliptic plane. A survey like Euclid, can expect to detect $1.4\times10^{5}$ solar system objects \citep{Carry2018}, high inclination (i $>$ 15$^\circ$) asteroids \citep[for  which there is currently a 
bias against in current census, see][]{Mahlke2018} and possibly even some rare interstellar objects such as the recently discovered 1I/'Oumuamua \citep{Meech2017,Katz2018}. It's simultaneous measurements in both visible and near-infrared will enable us to detect and compositionally map SSOs at the same time. It will nicely complement from the visible photometry from LSST and spectroscopy from Gaia \citep{Delbo2012} in mapping the dynamics of SSOs. Identifying and removing asteroids in the data is also important for weak gravitational lensing to prevent contamination of the shear signal \citep{Hildebrandt2017}. Since Euclid will produce close to a terabyte of data per day, we need to prepare tools to deal with this big data quickly and accurately. 

Machine learning is ideal approach to tackle the large data volume and the speed required to deal with upcoming Euclid data. Machine learning and neural networks have been used in astronomy for several years: \cite{Odewahn1995} used such methods to classify the morphology types of galaxies from their properties, similarly \cite{Gulati1994} built a neural network to classify stellar spectra and on the topic of asteroids, \cite{Misra2008} trained a neural network to predict the spectral class of asteroids from SDSS\footnote{Sloan digital sky survey \url{https://www.sdss.org}} data. 
Furthermore, convolution neural networks (CNNs) have allowed us to apply machine learning directly on astronomical images:  \cite{Dieleman2015} used CNNs to classify galaxy morphologies, \cite{Schaefer2018} to detect strong gravitational lensing and they have even been used to estimate continuous properties such as photometric redshifts \citep{Pasquet2019} and galaxy cluster masses \citep{Ntampaka2018}. 

In the present study we apply machine learning techniques to the problem of solar system object (SSO) detection. The paper is structured as follows: in \autoref{sec:data} we describe the data and simulations, in \autoref{sec:method} we present the convolutional neural net architectures, \autoref{sec:results} describes our results and we conclude in \autoref{sec:conclusions}.


\section{Data}
\label{sec:data}
Euclid\footnote{\url{http://sci.esa.int/euclid/}} is a 1.2m optical and near-infrared space telescope, that will observe $\sim15,000$ deg$^2$ of the sky (or over a third of the extragalactic sky) down to a $\rm V_{AB}$ magnitude $\sim24.5$ over its planned mission time of 6.25 years. The Euclid survey will be carried out using the step and stare technique: the 0.5 deg$^2$ field of view will observe the same portion of sky four times, using an optimised dither pattern \citep{Racca2016} and as a result of this planned mission strategy, it should be trivial to identify any moving object within the solar system. However with the presence of cosmic rays, galaxies and instrumental effects there is a lot of room for the mis-identification of SSOs. 

The Euclid payload consists of 2 instruments. VIS \citep{Cropper2016} will obtain high-resolution optical imaging and NISP \citep{Maciaszek2016} will provide photometry in three near-infrared bands as well as slit-less spectroscopy measurements. 

\subsection{Simulations}
\label{sec:simulations} 
In order to develop the science ground segment tools (pipeline, data analysis software, system infrastructure, etc.) in preparation for the Euclid launch and to assess its capabilities in meeting its scientific goals, detailed and extensive simulations of Euclid dataset are carried out within the Euclid consortium. Our simulations are created using instrument simulators developed for such purposes. The simulator we use is based on the Euclid Visible InStrument Python Package (VIS-PP)\footnote{\url{http://www.mssl.ucl.ac.uk/~smn2/}}. 

We ingest simulations of SSOs to create state of the art, Euclid-like images for training our model. The simulated images are generated as follows:
\begin{enumerate}
\item The simulator reads in a catalogue of objects (stars, galaxies, SSOs) with coordinates, magnitude, orientation and, for the SSOs, apparent speed.
\item For the objects that fall onto the CCD, the number of electrons are computed from the object's magnitude.
\item If the object is a galaxy it is simulated as an input snapshot taken from Hubble and then convolved with the Euclid PSF.
\item If the object is an SSO, this is simulated as a trail of aligned stars. An oversampling factor of 10 is used to avoid PSF under sampling effects (e.g. if an SSO covers 10 pixels then it will be generated from 100 stars). The position of each SSO star is determined by the input speed and orientation angle, whilst the apparent magnitude of the SSO is determined by the integrated stellar flux, 
\begin{equation}
	m_{*} = m_{\rm SSO} + 2.5\log_{10}\left( N_* \right).
\end{equation}
\item Once the noiseless image is generated, electron bleeding, ghosts, cosmic rays, charge transfer inefficiency, read-out and dark noise, conversion from electrons to ADU and additive bias are applied. We note that pointing inaccuracy and focal plane distortions are not simulated. As long as they can be fixed using the Gaia reference catalogue\citep{Gaia2018}, these effects should have no influence on the detection of SSOs. 

\end{enumerate}

\begin{figure} 
\centering
\frame{\includegraphics[width=2cm]{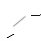}}
\frame{\includegraphics[width=2cm]{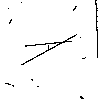}}
\frame{\includegraphics[width=2cm]{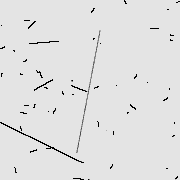}}
\frame{\includegraphics[width=2cm]{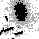}}
\subfloat[\tiny 20-21 mag, 10"h$^{-1}$]{\includegraphics[width=2cm, padding=1px]{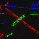}}
\subfloat[\tiny 20-21 mag, 40"h$^{-1}$]{\includegraphics[width=2cm, padding=1px]{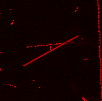}}
\subfloat[\tiny 20-21 mag, 80"h$^{-1}$]{\includegraphics[width=2cm, padding=1px]{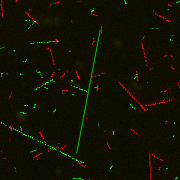}}
\subfloat[\tiny 25-26 mag, 10"h$^{-1}$]{\includegraphics[width=2cm, padding=1px]{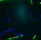}}
\caption{Examples of SSO images used in the single channel data set (top row) and those used in the three channel data set (bottom row). The SSOs are the objects closest to the centre of the image. The first three images are SSOs with magnitudes in the 20-21 mag bin and the last object is a 25-26 mag SSO. From left to right, the SSOs have speeds of 10, 40, 80 and 10 arcsec h$^{-1}$ respectively. The contrast levels of the images have been adjusted to enhance the appearance of object of interest, and all the images have been rescaled for illustration.  
\label{Fig: classes}}
\end{figure} 

\begin{figure} 
\centering
\frame{\includegraphics[width=2cm]{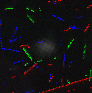}}
\frame{\includegraphics[width=2cm]{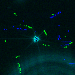}}
\frame{\includegraphics[width=2cm]{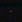}}

\subfloat[\tiny galaxy]{\includegraphics[width=2cm, padding=1px]{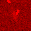}}
\subfloat[\tiny star]{\includegraphics[width=2cm, padding=1px]{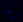}}
\subfloat[\tiny cosmic ray]{\includegraphics[width=2cm, padding=1px]{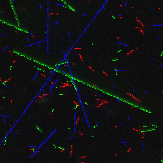}}
\caption{ Examples of the other classes used in the three channel data set. The left most images are galaxies, the centre two images are stars, and the right most images are cosmic rays. The lower left image has had adjustments made to the contrast levels to enhance the appearance of the galaxy, and all the images have been rescaled for illustration.
\label{Fig: classes3}}
\end{figure} 

\subsection{Preparing the data}
Our simulated images mimic the output of VIS - Each field of view consists of 4 quadrants. Each quadrant consists of a CCD readout node on the VIS instruments. The total size is 4096 x 4136 px, whilst each quadrant has a size 2048 x 2066 px. The image scale is 0.1 arcsec/px. Each field of view undergoes 4 dithering manoeuvres. The pointing displacements during the dithers have been optimised to the following, dither 2 - $\Delta$X: 100", $\Delta$Y: 50", dither 3 - $\Delta$X: 100", $\Delta$Y: 0", dither 4 - $\Delta$X: 100", $\Delta$Y: 0". Each dither slew takes 64 seconds depending on the altitude and orbit, and 280s for a field-to-field slew. We extract postage stamp cutouts of objects of 4 categories; SSOs, cosmic rays, galaxies and stars. The size of the postage stamp is chosen such that the image is filled by the object plus an additional padding value drawn from a gaussian distribution with mean 65 pixels and a standard deviation 5 pixels.
We investigate how our method performs both with (multi-channel) and without (single channel) temporal information from the dithers. In the multi-channel model, we combine dithers 1, 2, 3 and dithers 2, 3, 4 into 2 images of RGB channels and extract postage stamp cutouts centred on each object. We focus on the use of 3 out of the 4 Euclid dithers at a time due to constraints from our model (see \autoref{sec:transferlearning} for more details), but also show that it is possible to make use of all 4 dithers (\autoref{sec: Other}). For these images the pixels of galaxies and stars should appear in all 3 channels, whereas the pixels of cosmic rays and asteroids (aside from the very slow moving) should only appear in one channel. The differentiating characteristic feature of cosmic rays compared to SSOs are that the latter are convolved with the PSF. Examples of the postage stamps are shown in \autoref{Fig: classes} and \autoref{Fig: classes3}.


\section{Method}\label{sec:method}
\subsection{Neural networks}
\begin{figure*} 
\centering
\includegraphics[width=7cm]{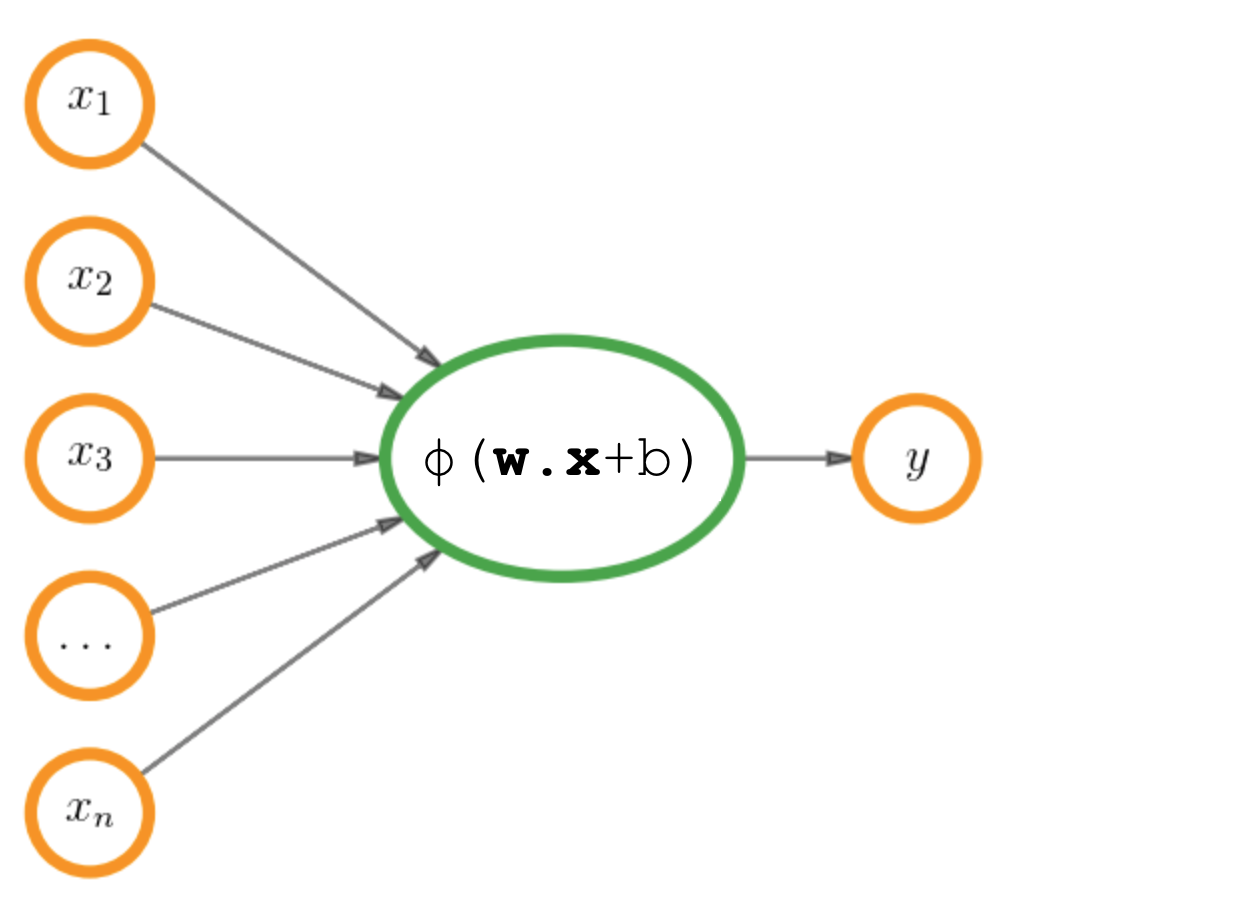}
\includegraphics[width=6.5cm]{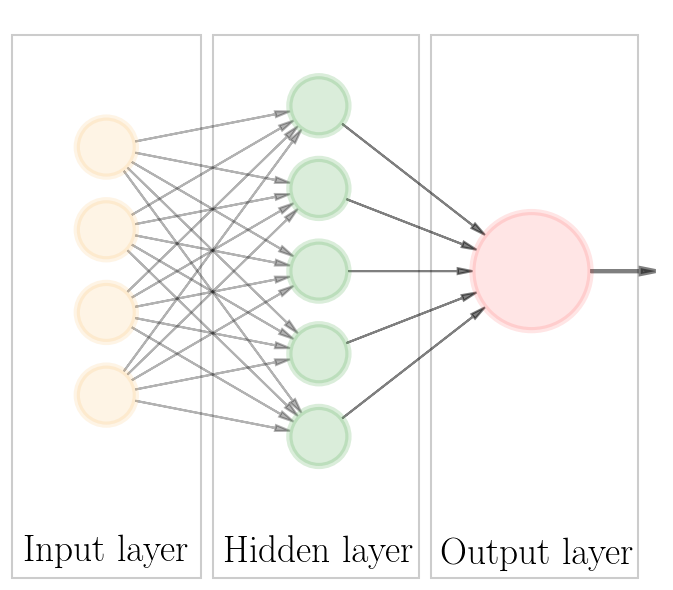}
\caption{\emph{Left:} Visualisation of a single neuron. \emph{Right:} Schematic diagram of a artificial neural network with one hidden layer. Each node in the hidden layer represents a neuron. A deep neural network will have multiple hidden layers. \label{Fig: Neuron}}
\end{figure*}

Artificial neural networks (ANNs) are a class of machine learning algorithm that maps some arbitrary inputs to outputs \citep{Mcculloch1943}. They were inspired by the visualisation process of the human brain and the hierarchical perception of images. In artificial neural networks, each neuron takes a vector of inputs $\bm{x} = \{x_1, x_2, ..., x_n\}$ and applies a set of weights $\bm{w} = \{w_1, w_2,..., w_n\}$ and a bias $b$ to it,
\begin{equation}
y = \phi \left(\sum_i w_i x_i + b\right).
\end{equation}
 The input is passed to every neuron in a layer and the output of each neuron will be passed as an input to each neuron in the subsequent layer (\autoref{Fig: Neuron}). The weights and bias values are parameters of the network, and $\phi(x)$ is a non-linear activation function which determines whether or not the neuron is fired, in other words it maps the input to the response. The most commonly used non-linear activation function is the rectified linear unit (ReLU), which takes the form, $\phi(x) = \max(0,x)$. This activation function is popular since both the function itself and its derivative is quick to calculate, 
\begin{equation}
\frac{\partial\phi(x)}{\partial x}= 
  \begin{cases} 
   0 &  x \leq 0 \\
   1       & x > 0
  \end{cases}.
\end{equation}

Furthermore, unlike sigmoid or tanh activation functions, ReLU saturates only half of the time so partially solves the vanishing gradient problem.

The input ($\bm x$) and parameters (${\bm w}$, $b$) give a predicted output $\hat{{\bm y}} = \{\hat{y}_1, \hat{y}_2, ..., \hat{y}_n\}$. In supervised learning the true output ${\bm y} = \{y_1, y_2, ..., y_n\}$ is known and backpropagation \citep{Rumelhart1986} is typically used to efficiently adjust the parameters to minimise a loss function $L({\bm y}, \hat{\bm y})$ (a measure of the distance between the true and predicted output). Using the gradient descent method to find the minima, the weights are iteratively updated,
\begin{eqnarray}
{\bm w}&\rightarrow& {\bm w} - \eta \frac{\partial L}{\partial {\bm w}}\nonumber \\
b &\rightarrow& b - \eta \frac{\partial L}{\partial b} 
\end{eqnarray}
Here $\eta$ is a tunable learning rate that determines the size of steps that the parameters can take on each update. A large learning rate will explore more of the parameter space but will make convergence to the minimum more difficult. A low learning rate is can be more precise but will take a long time to reach the minimum. In low dimensional parameters spaces there may be a risk of getting stuck in local minima if the learning rate is too small, however local minima are very rare in the high dimensional parameter spaces that are common to neural networks.

The calculation of the gradients can be computationally expensive when a cycle of the entire data set is required to update the parameters (batch gradient descent) and the number of training samples is large ($\mathcal{O}\sim10^6$). Alternatively the gradients can also be averaged over several randomly selected single samples or small samples (mini-batches) of the training data. This is known as stochastic gradient descent (SGD). It is much faster since it uses less RAM, and is better for finding multiple local-minima, however it requires the specification of the batch size. If the batch size is too small the convergence to the global minima will be slow and tends to be noisy \citep{LeCun2012}.
 
\subsection{Convolutional neural nets}\label{ssec:convnets}
\begin{figure*} 
\centering
\includegraphics[width=8.5cm]{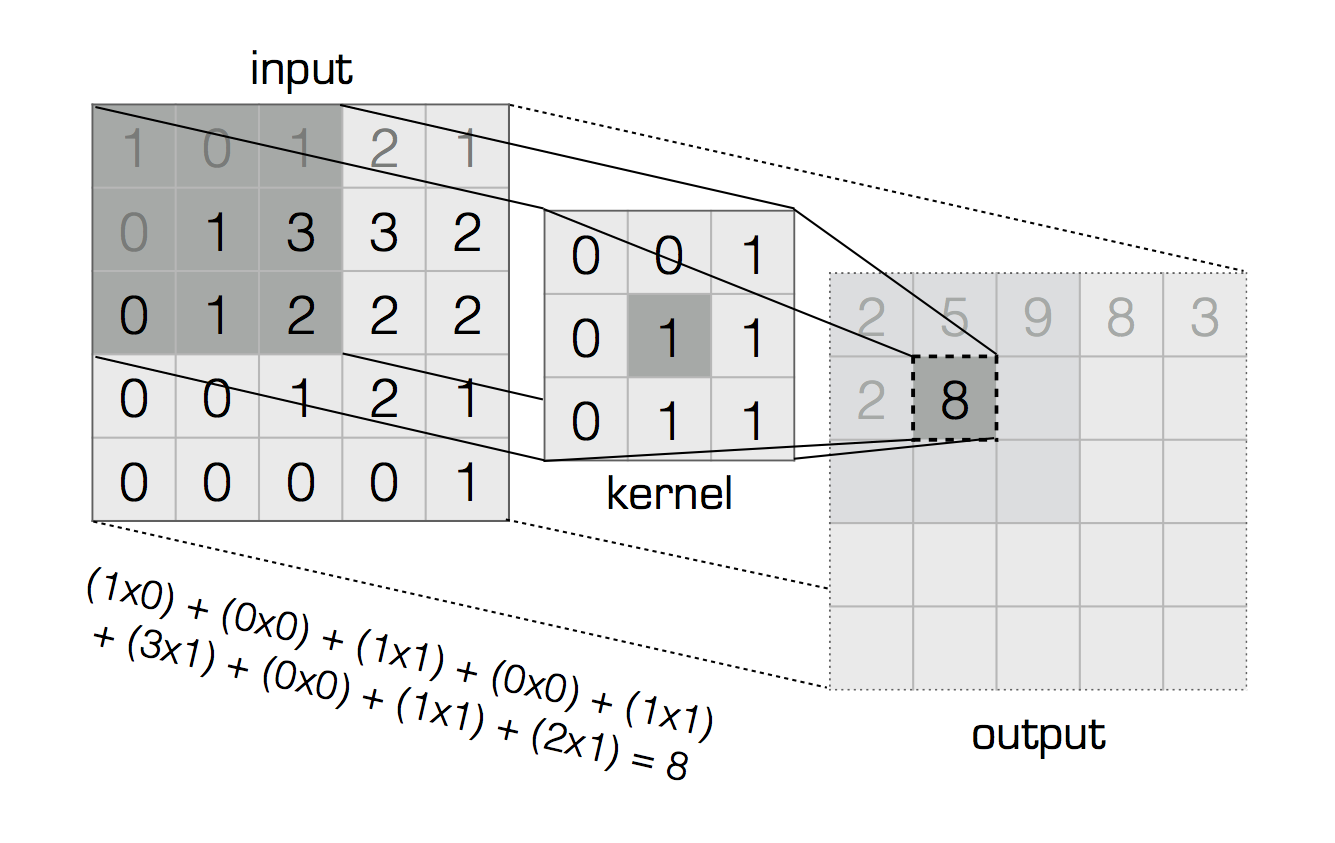}
\includegraphics[width=7cm]{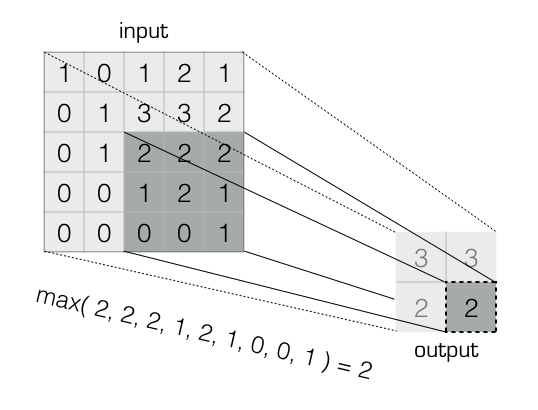}
\caption{\emph{Left:} Visualisation of a simple convolution filter layer with a 3x3 convolutional filter, no padding and a stride of 1. The input represents the pixel values of an image and the kernel is applied by sliding the kernel's centre pixel value over pixels of the input. The output pixel is a weighted sum of overlapping input and kernel pixels. \emph{Right:} Visualisation of a pooling layer, with a 3x3 maxpool and a stride of 2. Again this is applied sliding across the input pixels. The pooling layer reduces the dimensionality of the input by mapping only the highest pixel values. 
\label{Fig: CNN}}
\end{figure*}

\begin{figure*} 
\centering
\includegraphics[width=10cm]{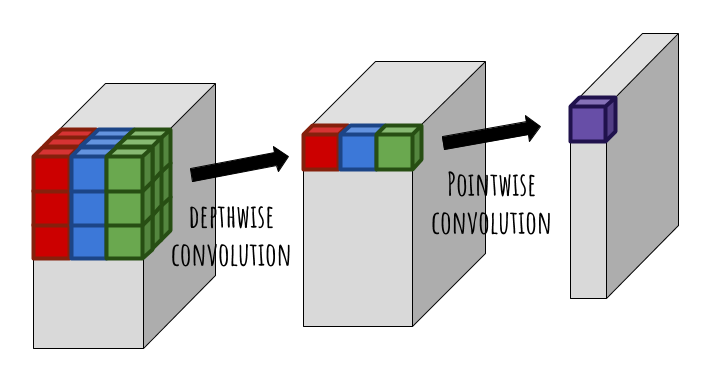}
\caption{ Depthwise separable convolution consist of first a depthwise convolution that is applied separately on each channel and then a pointwise convolution that is the same as a normal convolution (combines all channels) with a 1x1 kernel. 
\label{Fig: depthwise_seperable}}
\end{figure*}

Convolutional neural nets \citep[CNNs,][]{LeCun1998} are deep artificial neural networks designed for image recognition. Similarly, the CNN architecture contains multiple hidden layers, local connections between nodes and spatial invariance. The convolutional layer of a CNN takes an input image and convolves it with a small filter (kernel) whose values are weight parameters to be learnt. The filter is applied across the entire image with a bias and activation function creating a feature map layer.  \autoref{Fig: CNN} - left,  shows an example of a simplified CNN. For images where there are multiple channels (e.g RGB colour images), standard convolutional filters will combine all the channels using a weighted sum. Computationally this is not very efficient. Pooling layers are used to reduce the dimensionality of the feature maps, usually by reducing small areas to their maximum pixel value (\autoref{Fig: CNN} - right). An alternative to the standard convolutional filter is to use depthwise separable convolution (\cite{Chollet2016}, \autoref{Fig: depthwise_seperable}) which requires less parameters since the convolutional filters are first applied on each colour channel separately and then on each pixel across all channels. This also reduces the sensitivity to small positional shifts and distortions of a feature. Batch normalisation layers are used to improve speed and generalisation of the trained network by re-normalising mini-batches to their mean and variances during training, and fully connected layers connect all neurons in the previous layer and enables the mapping to a classification label. 

\begin{table}
\centering
\hspace*{-0.8cm}	\begin{tabular}{l | c | c | c | c |}
		Model 							& Number of 	& Top-1 			& Top-5 & input	\\
										& parameters	& error*  		& error* & size (px)\\
		\hline
		Inception v4 						& 35M 		& 80.2 			& 95.2 &	299$\times$299		\\
		MobileNet$\_$v1$\_$1.0$\_$224 	& 4.2M 		&70.7			& 89.5 & 224$\times$224			\\
		NASNet-A$\_$Large$\_$331 & 88.9M 		& 82.7			& 96.2 &331$\times$331			\\
	\end{tabular}
	\caption{CNN architecture versions used in this paper. *Taken from Google's internal training on the ILSVRC-2012-CLS\protect\footnotemark  dataset using single image crop and may differ from values listed else where. \label{Tab: architectures}}
\end{table}
\footnotetext{\url{http://www.image-net.org/challenges/LSVRC/2012/}}

\begin{figure*} 
\centering
\includegraphics[width=2cm, valign=c]{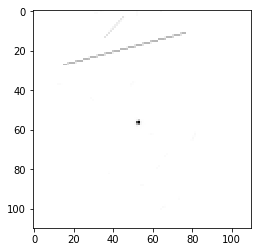}
 $\Longrightarrow$
\includegraphics[width=2cm, valign=c]{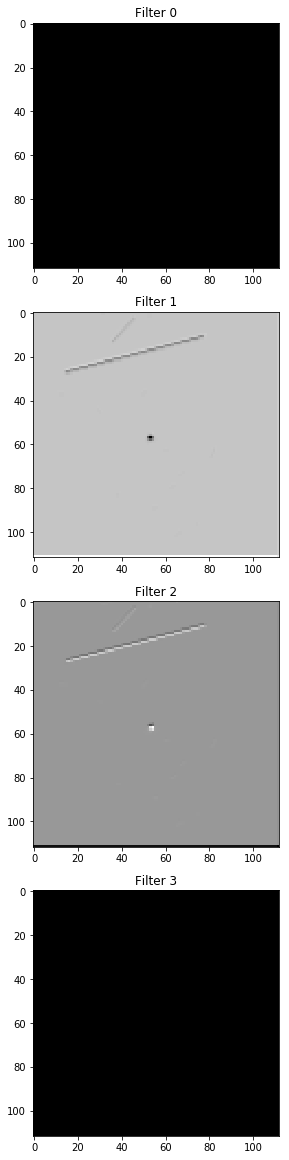}
 $\Longrightarrow$
\includegraphics[width=2cm, valign=c]{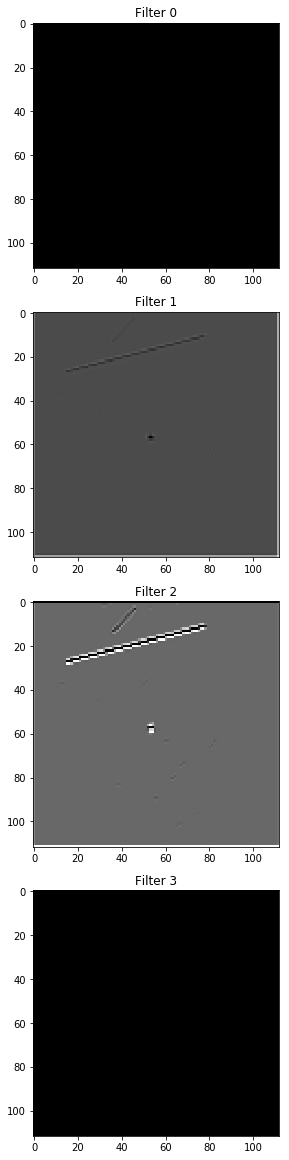}
 $\Longrightarrow$
\includegraphics[width=2cm, valign=c]{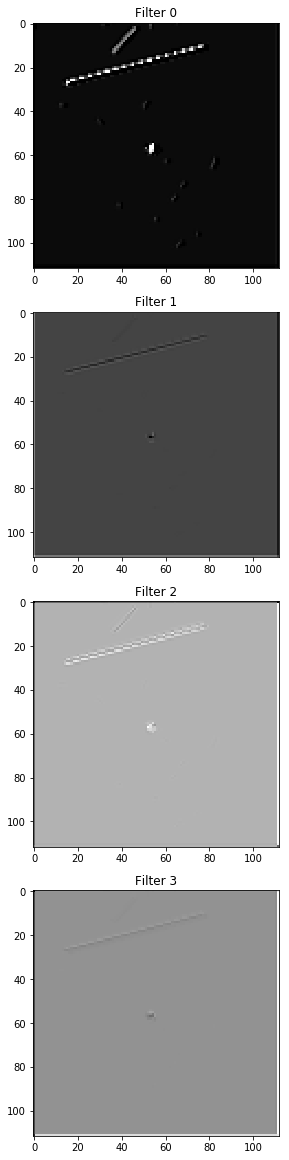}
 $\Longrightarrow$
\includegraphics[width=2cm, valign=c]{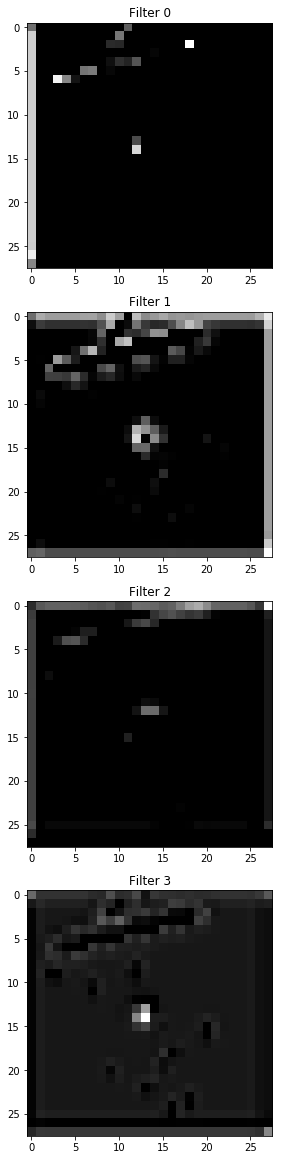}
 $\Longrightarrow$
\includegraphics[width=2cm, valign=c]{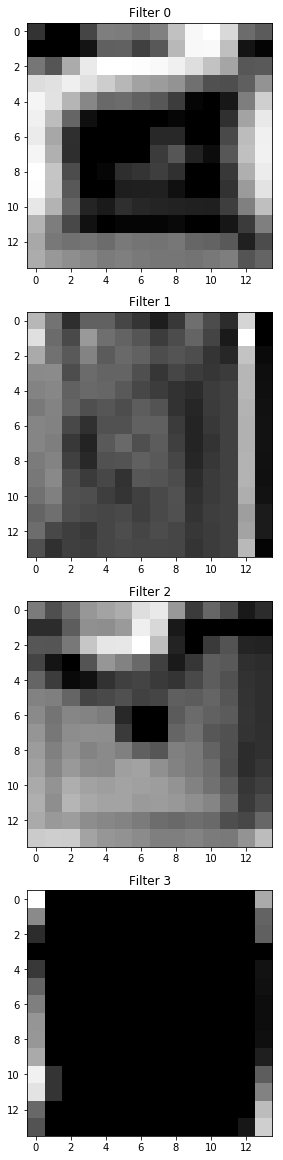}

\bigskip

 \makebox[2cm]{Input}
  \phantom{$\Longrightarrow$}
 \makebox[2cm]{Layer 1} 
   \phantom{$\Longrightarrow$}
 \makebox[2cm]{Layer 2}
  \phantom{$\Longrightarrow$}
 \makebox[2cm]{Layer 4}
   \phantom{$\Longrightarrow$}
 \makebox[2cm]{layer 11}
  \phantom{$\Longrightarrow$}
 \makebox[2cm]{layer 14}
\caption{ Activation images showing an input image (here a star) changing after convolution with the mbnet filters. Here we show only the outputs of the first 4 filters of each layer. Layer 1 contains 32 3$\times$3$\times$3 standard convolutional filters,  layer 2 contains 32 3$\times$3 depthwise filters, layer 4 contains 64 1$\times$1$\times$32 pointwise filters, layer 11 contains 256 3$\times$3 depthwise filters and layer 14 contains 512 3$\times$3 depthwise filters. Also note that the image size decreases with the layers due to the pooling layers.
\label{Fig: activations}}
\end{figure*} 

\subsection{Architectures}\label{sec: arch}
The layers of a neural network make up its architecture. Designing a good architecture is difficult, time consuming and requires expert knowledge. The hyper-parameters, those that need to be defined before training, include the number of convolutional filters, filter size (typically 3x3 pixels), padding (which defines the margin used when the convolutions are applied), stride (which determines if any pixels are skipped during the convolution), pooling layer size and dropout (a random probability of a neuron being ignored) to name a few. Whilst much research has been done to choose the best hyper-parameters \citep[e.g.][]{Simonyan2015, Szegedy2015, Murugan2017}, in practice it is trial and error, and complexity is added slowly. We use Google's Tensorflow library\footnote{\url{https://www.tensorflow.org}} and three different architecture models (\autoref{Tab: architectures}). 

Inception \citep{Szegedy2015} is a state of the art convolutional neural network. The performance is often defined by the percentage of correct classifications featuring in the top-1 and top-5 predicted categories of each image. In 2015, \texttt{Inception$\_$v3} was the first CNN to bypass the average human top-5 error rate of 5$\%$, obtaining a top-5 error of 3.5$\%$ on the ILSVRC-2012 dataset of 1000 categories. The original \texttt{Inception-v1} deep convolutional architecture was named \texttt{GoogLeNet}. It was inspired by \cite{Lin2013}'s \texttt{Network In Network}, and was one of the first CNN architectures to implement modules of parallel operations (inception modules) instead of the traditional sequential stacking. We implement the latest version \texttt{Inception$\_$v4} \citep{Szegedy2016} which has since been refined to include batch-normalisation, additional factorisation ideas, more inception modules and a more simplified uniform architecture. 

Another CNN we use is \texttt{NASNet-A-large} \citep{Zoph2017}. Nasnet is an architecture that was created by a controller neural network \texttt{Neural Architecture Search (NAS)}. With a small dataset, \texttt{NAS} uses a recurrent neural network \citep[RNN,][]{Zoph2016} and reinforcement learning to continuously propose improvements to the architecture. Since the use of \texttt{NAS} on larger datasets would be computationally expensive, it was applied to CIFAR-10, a dataset of 8$\times10^7$ small images in 10 classes to search for an optimal but scalable convolutional cell.  Using a larger dataset Imagenet 2012\footnote{\url{http://image-net.org/}} which consists of $\sim$ 1.4$\times 10^7$ images and 1000 classes, \texttt{NASNet-A-large} retrained the weights of its on repetitions of these cells and filters in the penultimate layer. The final architecture consists of 18 repeated cells with 168 convolutional filters per cell. \texttt{NASNet-a-large} exceeds the performance of any human-designed model to date. 

The last architecture we use is \texttt{mobilenets$\_$1.0$\_$224} \citep[mbnet,][]{Howard2017} which is the most accurate of the MobileNets architectures. MobileNets are not conventional CNN's. Traditional mobile neural networks relied on cloud computing, but the increasing computational power of mobile devices means that we no longer require dependency on an internet connection to perform deep learning tasks. Whilst other new CNN's focus on maximising accuracy, MobileNets were designed to be very small and very fast. The standard convolutional filters in CNN's are factorised into depthwise and pointwise (1$\times$1) convolutional filters which allow MobileNets to achieve competitive classification accuracies whilst optimising for latency, size, and power restrictions. \autoref{Fig: activations} is an example of how the layers of the MobileNets architecture affect an input image.

\subsection{Transfer learning}\label{sec:transferlearning}
Deep neural networks can have millions of parameters that can take weeks, if not months to train, transfer learning \citep{Donahue2013} significantly reduces the time required for training without requiring GPU. In transfer learning there is no need to fully train a deep architecture. An existing architecture can be retrained and fine tuned to new classification labels \citep[see e.g. ][]{Khosravi2018}. This method has already been successfully applied in astronomy to detect galaxy mergers \citep{Ackermann2018} and to classify galaxy morphologies \citep{Sanchez2018}. For Euclid the expected number of SSO detections is very small at high elliptical latitudes (a few per field of view) but as many as thousands on the ecliptic galactic fields. Our simulated data are generated to be representative of the expected abundances of galaxies, stars and cosmic rays, but we use an asteroid abundance of 0.6 arcmin$^{-2}$. The simulations rely on Hubble data and are both computationally expensive and volume heavy to generate, therefore in order to have a balanced dataset of each class, we are restricted to a small dataset ($\mathcal{O}\sim10^3$). Consequently we implement transfer learning to avoid overfitting and to prevent getting stuck at local minima. The layers of weights are already defined through pre-training on the ImageNet 2012 dataset. We only need to retrain a new top layer to include the new classes of images and to preprocess the input images to conform to the input of the architectures (see \autoref{Tab: architectures}). For preprocessing we resize the postage stamps to the input size of the the architecture using bilinear interpolation. We further renormalise the image by subtracting all pixel values by 128 and dividing by 128.

CNN layers consist of a 3 dimensional volume of neurons with a width, height and depth. The depth means that we can incorporate the dithers of Euclid, with each dither corresponding to a different depth layer. Since the architectures we use were pre-trained on RGB (3 channel) images, we use only 3 of the 4 dithers at any one image. 

For retraining we append a new top layer that consists of a softmax activation function,
\begin{equation}
\phi(x_i) = \frac{\exp(x_i)}{\sum_{n=1}^N \exp(x_n)}
\end{equation}
that gives us a probability of each class ($i$) over all ($N$) classes, and a fully-connected (dense) layer. Since we use the softmax function, the output of our models are probabilities assigned to each class label. We take the class label with the highest probability as the predicted class. For optimisation we use the standard gradient descent (\texttt{GradientDescentOptimizer}) and minimise on the mean cross entropy loss (\texttt{softmax$\_$cross$\_$entropy$\_$with$\_$logits$\_$v2}),
\begin{equation}
L({\bm y}, \hat{\bm y}) =  - \sum_i p_i \log q_i
\end{equation}
where $ \bm p \in \{ \bm y, 1- \bm y\}$ and $\bm q \in \{ \hat{\bm y}, 1-\hat{\bm y}\}$. We initiate the learning rate at 0.0001 and reduce it by 5$\%$ every 1000 iterations. This provides good accuracy within an acceptable training time.

\begin{figure*}
\centering
\includegraphics[width=8.5cm]{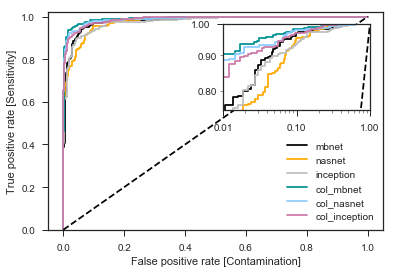}
\includegraphics[width=8.5cm]{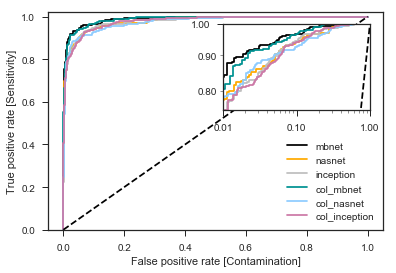}
\caption{ROC curve for the test data set with the same log scale plot inset. \emph{Left:} 2 category model. \emph{Right:} 4 category model. The dashed line represents a false positive rate equal to true positive rate. This would be equivalent to the model prediction occurring purely by chance. The colours indicate different architectures and data used.
\label{Fig: ROC}}
\end{figure*} 

\begin{figure*} 
\centering
Loss and accuracy curves for 2-label model. \vspace*{-0.5cm}

\includegraphics[width=7cm]{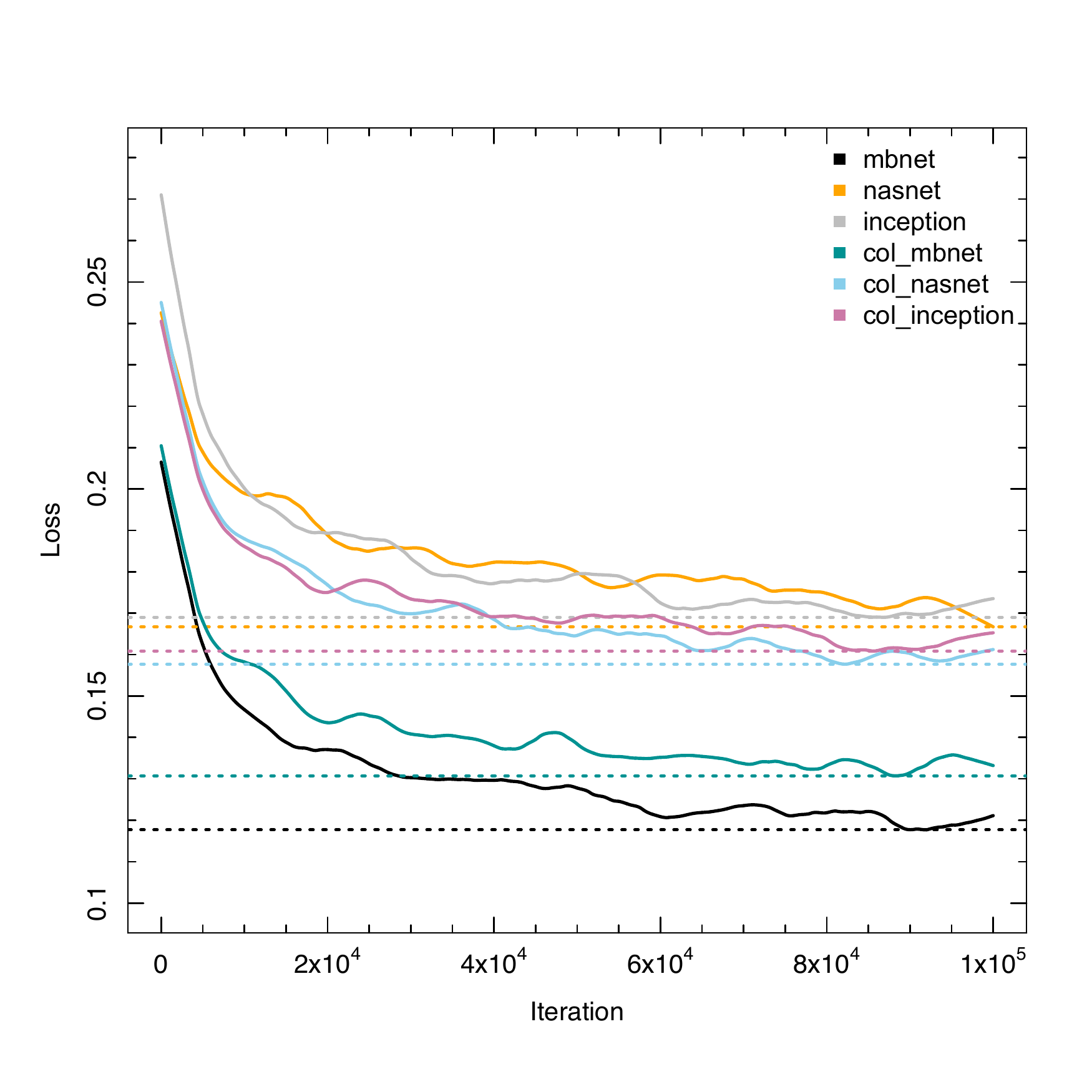}
\includegraphics[width=7cm]{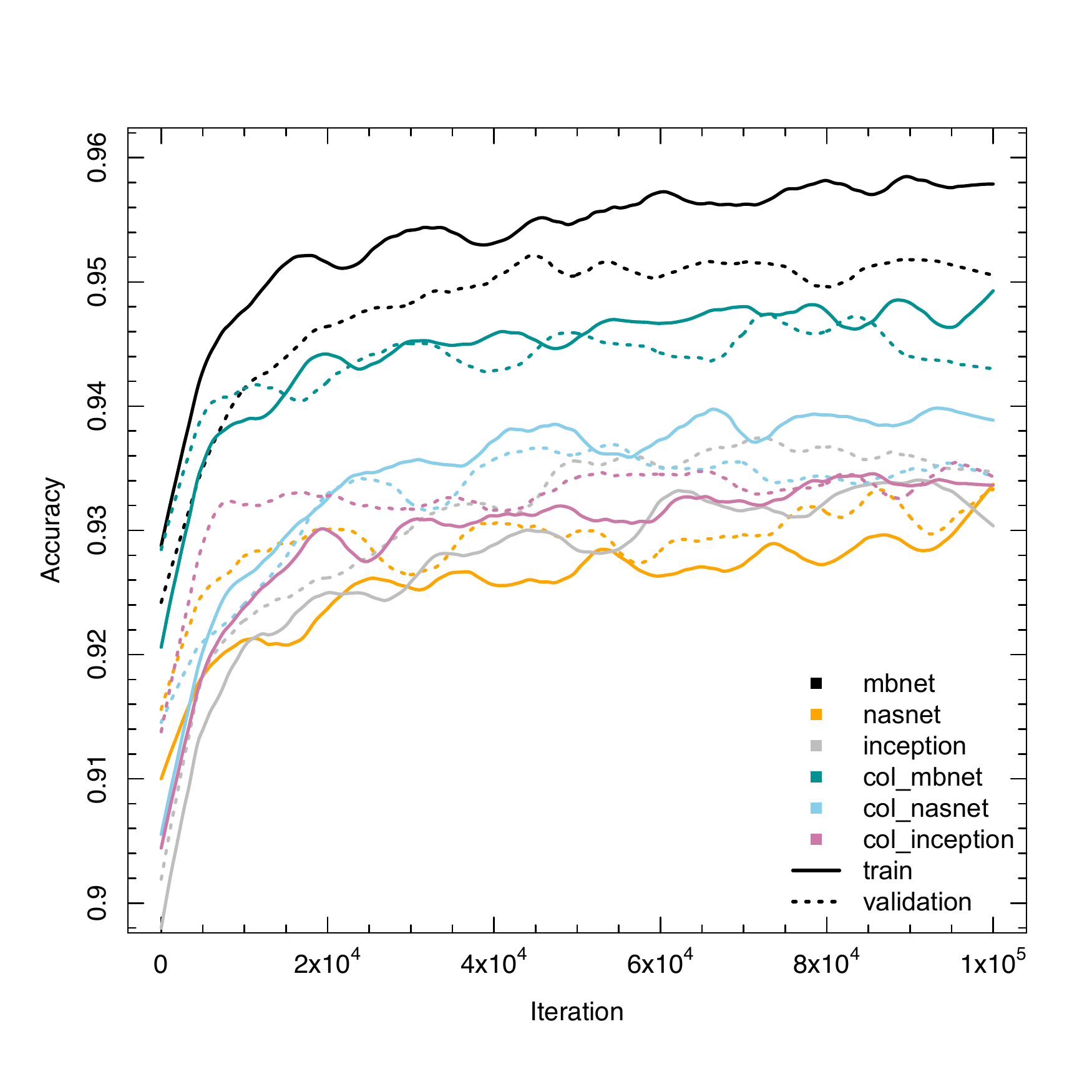}

Loss and accuracy curves for 4-label model. \vspace*{-0.5cm}

\includegraphics[width=7cm]{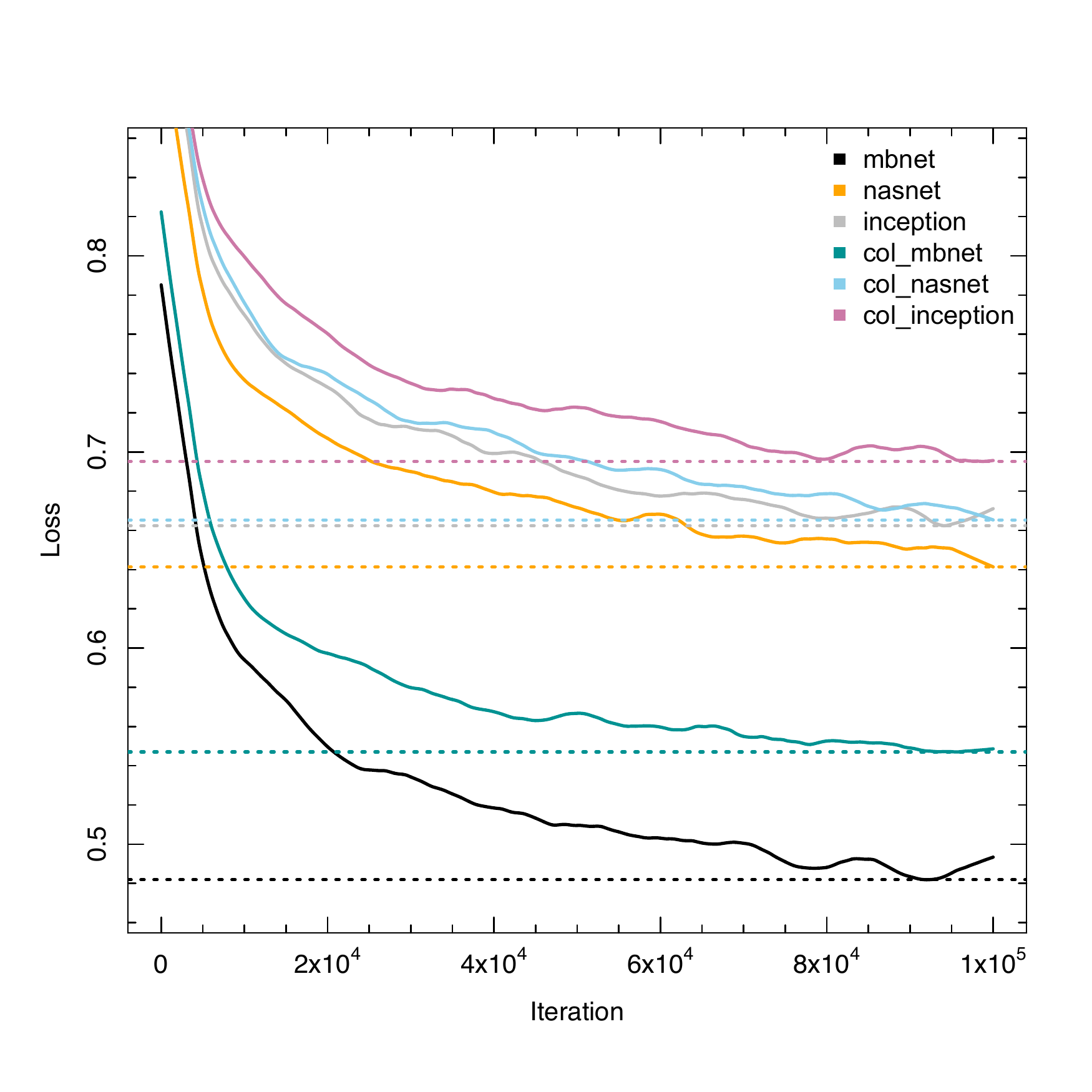}
\includegraphics[width=7cm]{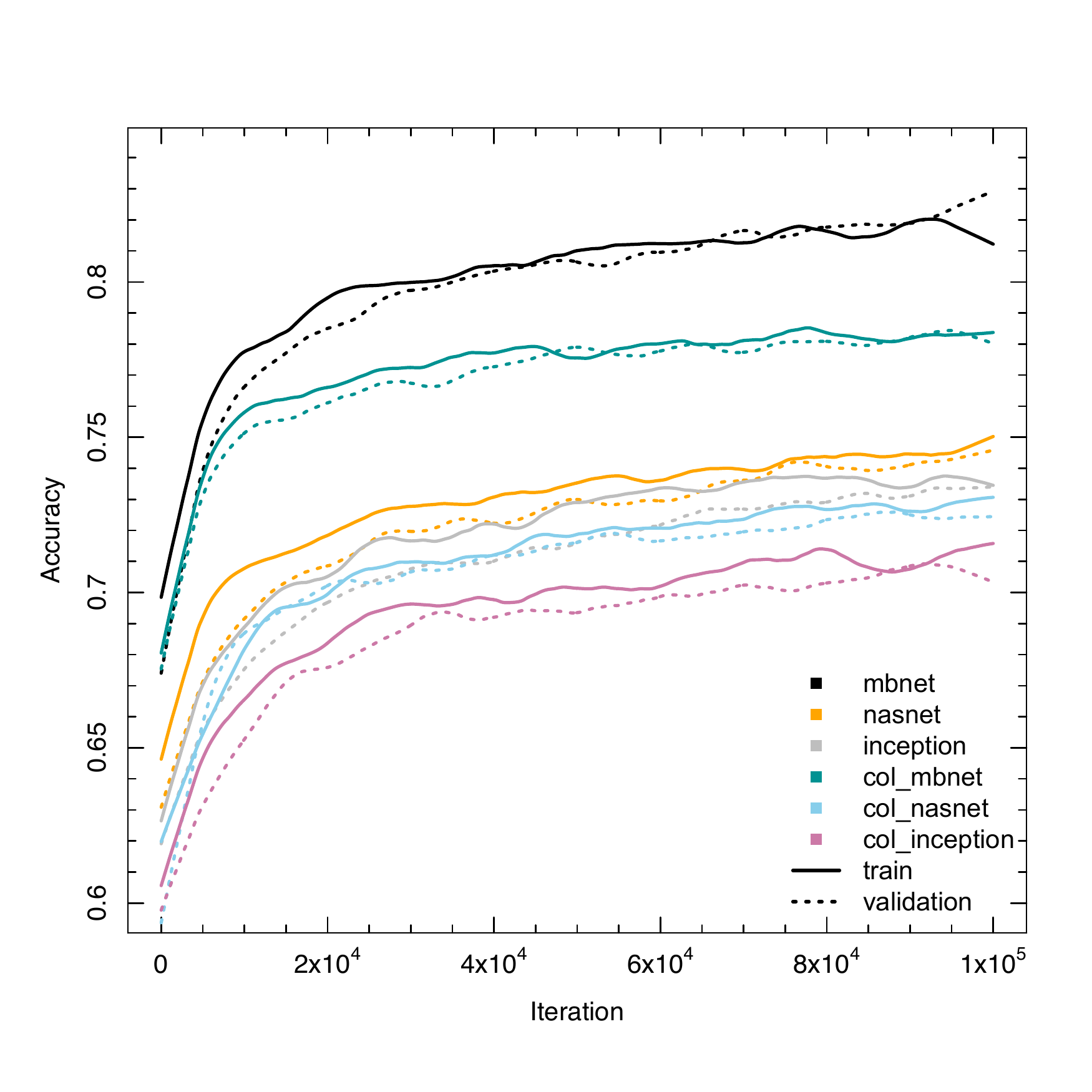}
\caption{ \emph{Top left: } Training loss as a function of iteration for 2-label runs. The dotted line shows the smallest loss and is solely for visual purposes. \emph{Top right:}  Training and validation accuracy as a function of iteration for the 2-label runs.\emph{Bottom:} the same as the top row but for the 4-label models.
\label{Fig: loss_convergence}}
\end{figure*}

\begin{figure*} 
\centering
Purity-completeness curves for an SSO abundance of 0.0001. 

\includegraphics[width=7cm]{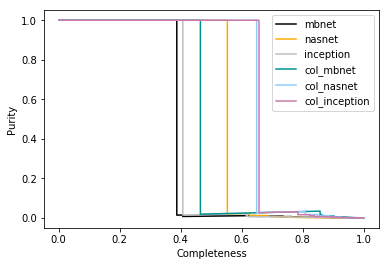}
\includegraphics[width=7cm]{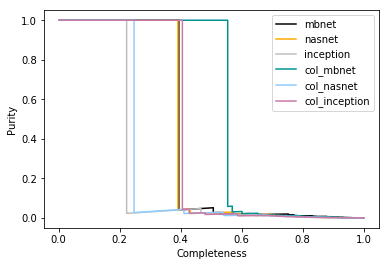}

Purity-completeness curves for an SSO abundance of 0.5.

\includegraphics[width=7cm]{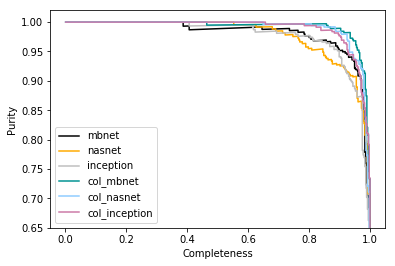}
\includegraphics[width=7cm]{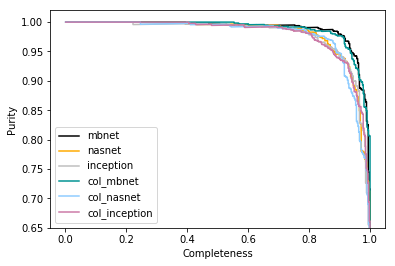}
\caption{ Purity - completeness curves. The plots on the left are the 2-label models and the plots on the right are the 4-label models. On the top row we use an SSO abundance of 0.0001 and on the bottom row an SSO abundance of 0.5.
\label{Fig: purity}}
\end{figure*} 
\subsection{Regularisation and augmentation}
Apart from transfer learning and increasing the amount of data, there are several other tricks that can be used to prevent overfitting. Regularisation by batch-norm was discussed in \autoref{ssec:convnets}, another regularisation technique is dropout, where connected nodes are randomly disconnected. We implement a 20$\%$ drop-out probability. Note however that the dropout does not significantly improve fits since there are few parameters in the final layer and convolutional neural networks are applied across several locations of an image. Augmentation prevents overfitting by adjusting the properties of the training data. Resizing, cropping, rotating, transposing, and other image adjustments can also improve the results however this also significantly increases the time required for training. In \autoref{sec: Other} we investigate the use of augmentation.

We split our data into training, validation and test sets with 70$\%$, 20$\%$ and 10$\%$ respectively. The training is run using Monte Carlo cross-validation \citep{Xu2001} and stochastic gradient descent with validation batches and training mini-batches of 100 images. The validation is performed every 10th iteration and the test data is only seen after the training is complete and is not used to update the parameters of the architecture.


\begin{table*}
\centering
	\begin{tabular}{l | c | c | c | c | c | c | c | c | l | c | c | c| c|}
		Model 							& time	& iterations 	& test  			& test 	& precision & recall  & F1 & cohen $\kappa$ & predicted& \multicolumn{2}{|c|}{truth} & &\\
										& (s)		&			& acc ($\%$) 	& AUC	 & & & score	     & \\
		\hline
		\multicolumn{10}{|l|}{2cat} 			& SSO 		& non-SSO\\
		\hline
		Inception 							& 9048 		&   100k		&  0.922		&  0.976	& 0.938 &	0.899	&0.918	& 0.843	& SSO 		& 320 & 21  \\
		 								& 			& 			& 			&		&	     & 			&		& 		& non-SSO 	& 36 & 350 \\
		MobileNet 						& 5485		&   100k		&  0.935		&  0.981	& 0.948 & 0.919 	& 0.933 	& 0.871 	& SSO 		& 327 & 18\\
		 								& 			& 			& 			&		&	     & 			&		&		& non-SSO 	& 29 & 353\\
		NASNet							& 54141		&   100k		&  0.911		&  0.977	& 0.927 & 0.888	& 0.907	&0.821	& SSO 		& 316 & 25 \\
										& 			& 			& 			& 		&	     &			&		&		& non-SSO 	& 40 & 346 \\
		\hline
		\multicolumn{10}{|l|}{col$\_$2cat} 			& 		& \\
		\hline		
		Inception 							& 91557		& 100k 		& 0.943		&0.989	&0.954  &0.925		&0.939	&0.885	& SSO 		&356& 17 \\
		 								& 			& 			& 			& 		&	     &			&		&		& non-SSO 	&29	& 399 \\
		MobileNet							& 5737		& 100k		&0.956		&0.992	&0.973  &0.935		&0.954	&0.912	& SSO 		&360& 10 \\
		 								& 			& 			&			& 		&	     &			&		&		& non-SSO 	&25 	& 406 \\
		NASNet							& 89981		& 100k		&0.945 		&0.989	&0.952  &0.932		&0.942	&0.89	&SSO 		&359&18  \\
		 								& 			& 			&			& 		&	     &			&		&		& non-SSO 	&26  & 398  \\
		\hline
		\multicolumn{10}{|l|}{4cat} & SSO  & CR &  galaxy & star   \\
		\hline
		Inception							& 18808		&100k  		& 0.737		& 0.918	&0.739	&0.737	&0.736	&0.649	& SSO 		& 304 & 11 & 6 & 27  \\
										& 			& 			& 			& 		&	     	&		&		&		& CR 		&  13 & 283 & 69 & 60 \\												 				& 			& 			& 			& 		&	     	&		&		&		& galaxy 		&  23 & 36 & 249 & 45 \\
										& 			& 			& 			& 		&	    	&		&		&		& star 		&  5 & 42 & 35 & 208\\
		MobileNet							& 5825		& 100k		&0.830		&0.962	&0.831	&0.83	&0.830 	& 0.773	&SSO 		& 313 & 6 & 6 & 10 \\
										& 			& 			& 			& 		&	     	&		&		&		& CR 		&  9    & 305 & 43 & 37  \\
										& 			& 			& 			& 		&	     	&		&		&		& galaxy 		&  17  & 20 & 297 & 33 \\
										& 			& 			& 			& 		&	     	&		&		&		& star		& 6 & 41 & 13 & 260  \\
		NASNet							& 38389		& 100k		&0.756 		&0.924	&0.759	&0.756	&0.756	&0.674	& SSO 		&  299 & 10 & 5 & 26   \\
										& 			& 			& 			& 		&	     	&		&		&		& CR 		& 18 & 281 & 72 & 50  \\
										& 			& 			& 			& 		&	     	&		&		&		& galaxy 		& 24  & 26 & 264 & 38  \\
										& 			& 			& 			& 		&	     	&		&		&		&	 star 		& 4  & 55 &18 & 226 \\
		\hline
		\multicolumn{8}{|l|}{col$\_$4cat} \\
		\hline
		Inception							& 11921  		& 100k  		&  0.726		& 0.911	& 0.730  	&0.726	&0.727	&0.634	&SSO 		& 348  	& 22 		& 5 		& 17   \\
										& 			& 			& 			& 		&	     	&		&		&		& CR 		& 20		& 249	& 80 		& 32  \\
										& 			& 			& 			& 		&	     	&		&		&		& galaxy 		& 23		& 58		& 235	& 69 	\\
										& 			& 			& 			& 		&	     	&		&		&		& star 		& 13		& 47		& 30		& 269  \\
		MobileNet							& 6152		&  100k		&  0.790		& 0.949	& 0.793    	& 0.790	& 0.791	&0.719	& SSO 		& 359	& 15		& 3		& 3   \\
										& 			& 			& 			& 		&	     	&		&		&		& CR 		& 17		& 286	& 66		& 29  \\
										& 			& 			& 			& 		&	     	&		&		&		& galaxy 		& 24		& 37		& 245	& 47  \\
										& 			& 			& 			& 		&	     	&		&		&		& star 		& 4		& 38		& 36 		& 308  \\
		NASNet							& 25761		& 100k		& 0.725		& 0.914	& 0.727    	& 0.725	& 0.726	& 0.633	&SSO 		& 352	& 26 		& 6 		&17   \\
										& 			& 			& 			& 		&	     	&		&		&		& CR 		& 18		& 255	& 77		&38 \\
										& 			& 			& 			& 		&	     	&		&		&		& galaxy 		&  23		& 49		& 220	&59  \\ 
										& 			& 			& 			& 		&	     	&		&		&		& star 		&  11		& 56		&47		&273  \\
		\hline
		\multicolumn{10}{|l|}{Other tests}			& 		& \\
		\hline		
		random							& 296628		& 100k 		& 0.504		& 0.500 	& 0.000  	& 0.0		&0.0		& 0.0	 	& SSO 		&390		& 384 \\
				 											& 			& 			& 			& 		&	     	&		&		&		& non-SSO 	&0		& 0\\
		scratch							& 1219744	& 100k 		& 0.851		& 0.916 	& 0.866 	& 0.828	&0.847	& 0.703 	& SSO 		&341		& 66 \\
				 											& 			& 			& 			& 		&	     	&		&		&		& non-SSO 	&49		& 318\\
		aug							& 307607		& 100k 		& 0.957		& 0.988	& 0.944 	& 0.971	&0.958	& 0.915	& SSO 		&368 	&11 \\
				 											& 			& 			& 			& 		&	     	&		&		&		& non-SSO 	&22		& 373\\	
		4channel						&  796537		& 100k 		& 0.900		& 0.966	& 0.880 	& 0.934	&0.906	& 0.800	& SSO 		&305 	&25 \\
				 											& 			& 			& 			& 		&	     	&		&		&		& non-SSO 	&48		& 353\\												
	\end{tabular}
	\caption{Summary statistics for the different runs. The right most columns are the confusion matrix. This shows the number of predictions for each class against the true label when the model was applied to the test dataset. \label{Tab: results}}
\end{table*}

\subsection{Performance metrics}\label{sec: metrics}
Our testing set is based on simulations so the number of images in each class can easily be defined, however we do not train our model on an astronomically representative training dataset. Classification models generally do not perform well when trained on imbalanced datasets. They tend to overfit the more abundant class. This leads us to The Accuracy Paradox - a predictive model with a given accuracy, may have greater predictive power than a model with higher accuracy. Accuracy is defined as,
\begin{equation}
\rm Accuracy = \frac{TP + TN}{ TP + FP + TN +FN},
\end{equation}
where TP, FP, TN, FN are true positives, false positives, true negatives and false negatives respectively. It is clear that for an imbalanced data set with 1 SSO and 99,999 non-SSOs a high accuracy can be achieved by predicting all images to be non-SSOs, however this model would be useless for our purpose (detection of SSOs). 
Receiver Operating Characteristic (ROC) curve's are a common way to visualise the performance of a binary classification model. It shows the true positive rate (TPR = TP / (TP + FN)) against the false positive rate  (FPR = FP/ (FP + TN)) respectively.

Whilst accuracy and area under the (ROC) curve (AUC) are the typically preferred performance summary statistics, it only reflects the underlying class distribution and is not very informative when the training classes are imbalanced. It is therefore also important to consider: precision, a measure of how pure our sample is (i.e. what fraction are SSOs), recall, a measure of how complete our sample is (i.e. what fraction of all SSOs are in the sample), and the F1-score which is a weighted mixture of the two. A more relevant statistic for an unbalanced dataset is the Cohen kappa score, a measure of the accuracy normalised by the imbalance of classes. Another common solution is to use a weighted loss function or feed weighted samples to the mini-batch. With transfer learning, we have sufficient data to train on a balanced training set.

In astronomy, to determine how well a classifier method performs we can look at the purity and completeness of the samples. In machine learning, purity and completeness are equivalent to precision and recall. It is clear that purity and completeness is a trade-off, the sample can be very pure if the threshold value of SSOs is very high however this will lead to a low number of classified SSOs, likewise, the sample can be very complete by classifying all images (SSOs and non SSOs) as SSOs. 

However to see how well the CNN performs on real astronomical data we need to rescale our test results to the astronomical abundance of the classes.

\begin{equation}
\rm Purity \equiv Precision = \frac{TP}{TP+FP} ,
\end{equation}
\begin{equation}
\rm Completeness \equiv Recall = TPR = \frac{TP}{TP + FN},
\end{equation}
\begin{equation}
\rm Astronomical\, Purity = \frac{TPR \times AB}{TPR \times AB + FPR \times (1 - AB)}.
\end{equation}

where AB is the astronomical abundance (the number of SSOs / the number of all objects).

\section{Results and Discussion}
\label{sec:results}
\subsection{Considering two categories}
Initially we treat each dither independently and combine cosmic rays, galaxies and stars collectively into the category non-asteroids. The complete dataset contains 3756 images of SSOs and non-SSOs. Of the 3 models tested; mbnet, nasnet and inception (see \autoref{sec: arch}), we surprisingly found that the mbnet architecture out performed the other 2 on the test data set. Next we apply the same architectures to the 3 channel images (col$\_$mbnet, col$\_$nasnet, col$\_$inception) and similarly the top performing architecture is mbnet despite being the quickest architecture to train. We believe this is because our dataset consists of low resolution images and mbnet having been pre-trained on low resolution (224x224 pixel images) is able to better characterise low resolution features, whereas Inception and NASNet are both higher resolution (299x299 and 331x331 respectively) and deeper models but the added complexity is not adding any extra information and conversely suppressing the ability to generalise classification performance on new images. \autoref{Fig: ROC} shows the ROC curve for these models as applied to the test data set. The dashed line shows the expected ROC curve if the predictions from the model are down to chance. ROC curves above the dashed line perform better than a random guess and a ROC curve below the line would be performing worse than randomly guessing and is usually an indiction of a bug. As expected, adding the dither information improves the performance of the neural networks. 

\subsection{Considering four categories}
It is not uncommon within the astronomical community to write algorithms specifically to classify a single object, however convolutional neural nets, unlike other machine learning methods, do not require feature engineering which means the same network can be easily adapted to multiple class problems. It is therefore a more efficient use of personnel and resources to develop a single network that can identify multiple classes in the Euclid data than to have several networks each classifying a different astronomical object. We retrain the network with 4 labels; cosmic rays, galaxies, stars and SSOs, using single and multi-channel images (\autoref{Fig: ROC}). Each class contains 3756 images. The 4-label models perform slightly worse than 2-label model, this is not unexpected since the 2 label case requires at least a threshold probabilty of 0.51 to be classified as an asteroid, whereas the in the 4-label case it could be as low as 0.26. Also there are more degeneracies between classes. Once again the MobileNet architecture is the superior model.  \autoref{Tab: results} lists the performances of all the variations of architectures and schemes we run. From the confusion matrices, we find that stars are most often misidentified as SSOs in the network trained on single dither images, whereas, when we include the dither information this changes to cosmic rays. False negative SSOs were most often misclassified as galaxies in both scenarios.

\begin{figure} 
\centering
\includegraphics[width=8cm]{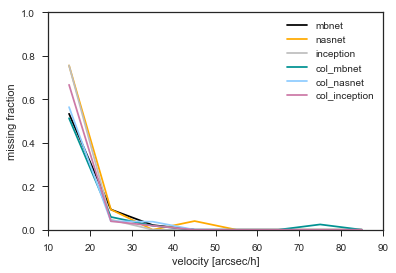}
\includegraphics[width=8cm]{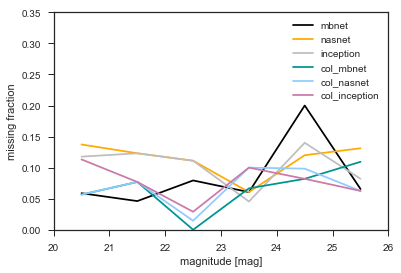}
\caption{ The fraction of SSOs incorrectly classified as a function of velocity and AB magnitude. Here we show results for the 2-label models however a similar consensus is drawn from the 4-label models. 
\label{Fig: missing}}
\end{figure} 

\subsection{Convergence}
The chosen number of iterations used in training needs to be carefully defined to avoid under or overfitting. This is when the model has not had enough time to learn the feature or has had too much time that it is completely fine tuned to perform well on the training set but performs badly on the test set. We monitor the training error as a function of the iteration to determine an ideal stopping point (\autoref{Fig: loss_convergence}). It also helps to determine a suitable learning rate. Ideally, the loss as a function of time will gradually decrease. If the decline is too fast, then the learning rate is too high and if the decline is too slow then the learning rate is too low. When the loss reaches a plateau, is a good place to stop training because if the training is run for too long it will overfit. If this is the case, the loss-iteration curve will start to upturn at large times. We also monitor the training and validation accuracies over time. These two curves should closely match each other if the neural net is not under or over fitting. 

\subsection{Purity and Completeness}
The expected abundance of asteroids observed with respect to the other classes in Euclid is as low as $\sim$0.0001 but varies with ecliptic latitude. For this abundance we can hope to achieve 100$\%$ purity for at most a 60$\%$ complete sample of asteroids (\autoref{Fig: purity}). None the less, it is clear that this significantly improves if the abundance of asteroids fed to the model is 0.5.  The neural network approach would make a good follow up method for the confirmation of potential asteroids detected from existing methods to improve the abundance ratio. Otherwise, another way purity can be further improved is to check the 4 dithers for counter part images \citep[see e.g.][]{Bouy2013,Mahlke2018}. We note however that an abundance of 0.0001 is a pessimistic estimate and an advanced Euclid pipeline would be able to remove the majority of cosmic rays.

\subsection{Biases}
To infer whether our method is biased in any particular way we look at the distribution of the false negatives as a function of asteroid AB magnitude and speed (\autoref{Fig: missing}). We find that there is no particular trend with magnitude, which means the CNN can pick up even the faintest asteroids rather well however it does not perform very well in picking out the slowest moving asteroids. From the confusion matrices of the 4-label models (\autoref{Tab: results}), we see that SSOs are most likely to be misidentified as galaxies which is not surprising since they are small, extended and convolved with the PSF just like the asteroids.

\begin{figure} 
\centering
\includegraphics[width=8cm]{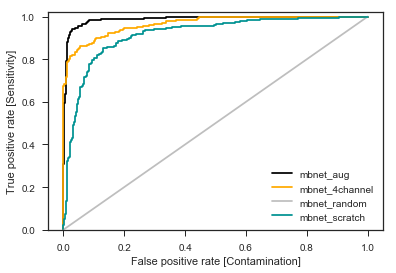}
\includegraphics[width=8cm]{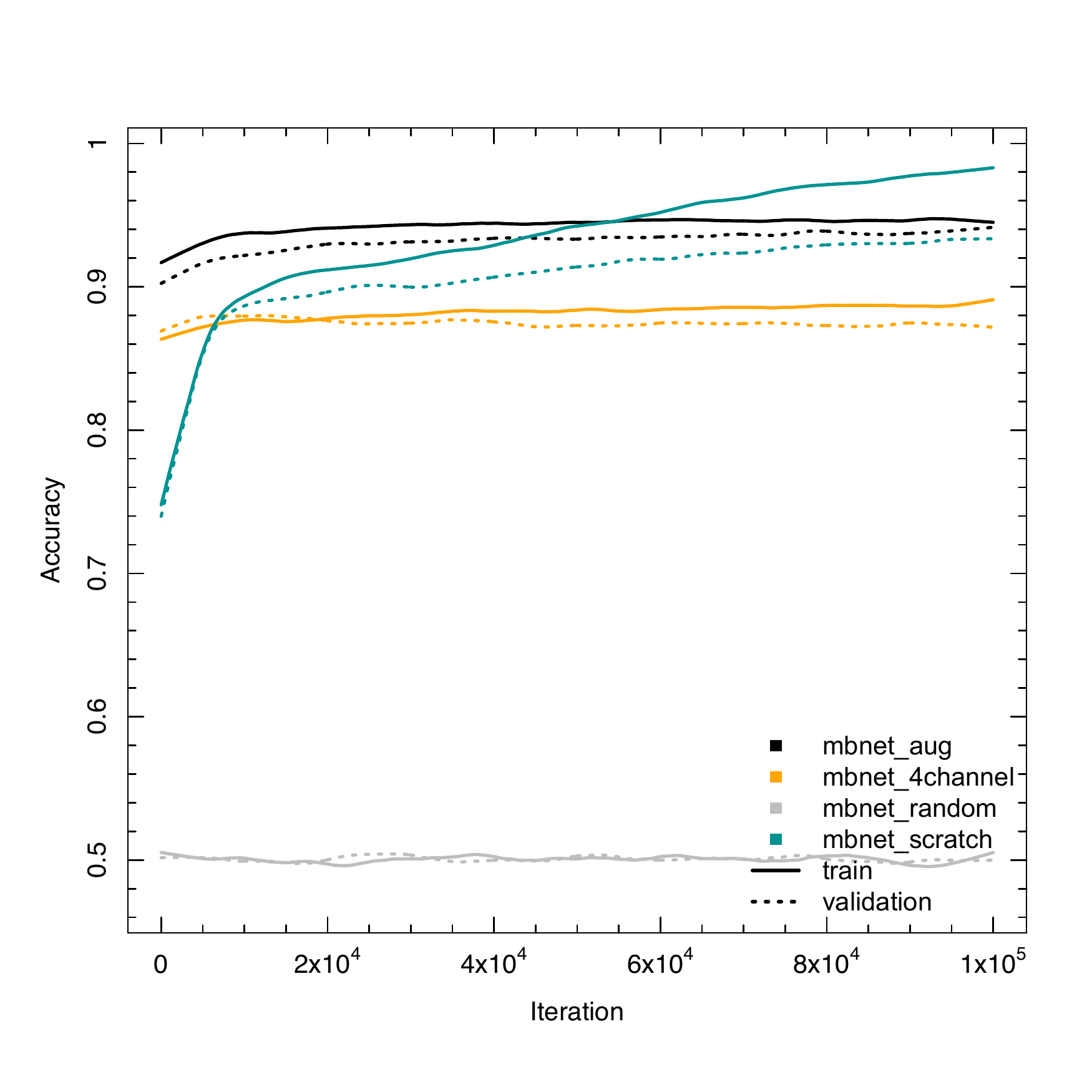}
\caption{Further architecture tests. The top plot shows the ROC curve of the test data for the networks using augmentation, 4 dithers, random weights and training from scratch. The bottom plot shows the accuracy of the training and validation as a function of iteration for the same networks.
\label{Fig: other}}
\end{figure}

\subsection{Further testing}\label{sec: Other}
We implement further testing on a cloud virtual machine. The following networks were trained on the 2-category scenario and the mobilenets architecture with single depth channel as the baseline. Whereas previously we used bottleneck values to reduce the computational time, here we do not. Firstly, to ensure the contribution from transfer learning is indeed significant, we re-run the training without loading the Imagenet pre-trained weights, instead randomising the weight values in the MobileNet architecture and fixing them (mbnet$\_$random). As expected we find the classification performance to be consistent with chance. The network classified all the test images as SSOs.

Furthermore we experimented training the MobileNet architecture weight values from scratch (mbnet$\_$scratch) with our data. On the training data and validation data this model seemed to perform the best, however on the test data it achieved 11$\%$ lower than the baseline model. On further examination of the loss and accuracy plots, we see that the model begins to overfit after 10,000 iterations. We believe this is due to the small data size that is insufficient to constrain the large number of parameter.

Augmentation of the training data is known to improve the robustness and performance of the model by effectively increasing the training data volume. We train a model with random flip, random image crop of up to 10$\%$, random image scaling of up to $\pm$ 10$\%$ and random brightness of up to $\pm$ 10$\%$. The augmentation improves the accuracy by 2.2$\%$ and the AUC by 0.7$\%$, but increases the training time by $\sim \times$50. Including augmentation meant that it was not possible to use bottleneck tensors of the pre-trained network output which significantly reduces the training time. Nonetheless, after training the neural network runs instantaneously, therefore if the computational power is not a concern then including augmentation is highly recommended.

Lastly, while it is not possible to use all four dithers as channels in the pre-trained architecture without training it from scratch, it is possible to use 4 dithers by modifying the architecture. To do this we must append additional layers before the pre-trained network, that reduce the dimensionality of the input data into the correct input dimensions for the pre-trained network. We use a 2D convolutional layer with a 3x3 kernel and 3 filters. This gave an accuracy and AUC of 4$\%$ lower and 1$\%$ lower than that of the baseline model, and 6$\%$ and 2$\%$ lower respectively compared to the 3 channel MobileNet model.

\subsection{Velocity predictions}\label{sec: Vel}
\begin{figure*} 
\centering
\includegraphics[width=8cm]{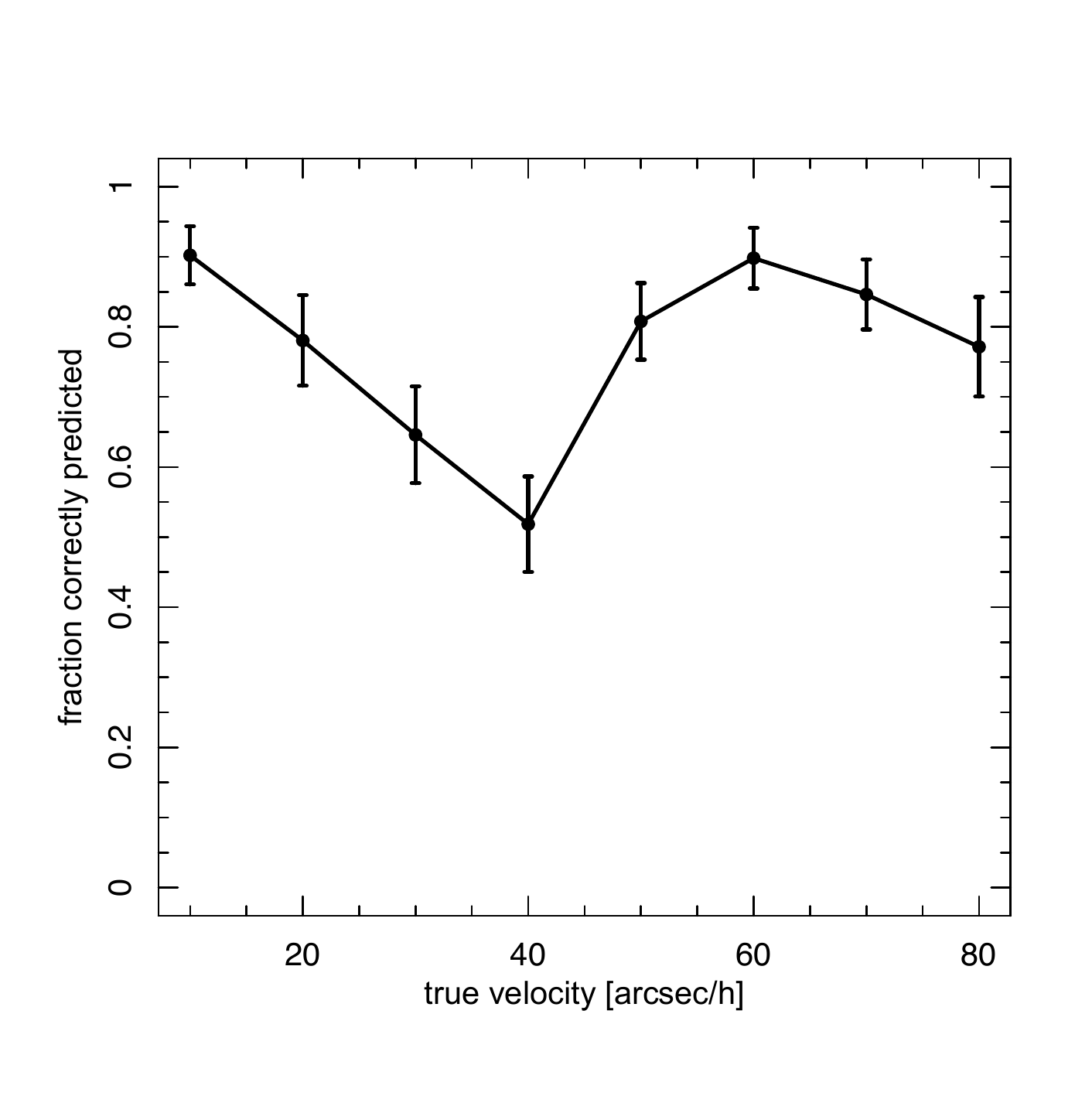}
\includegraphics[width=8cm]{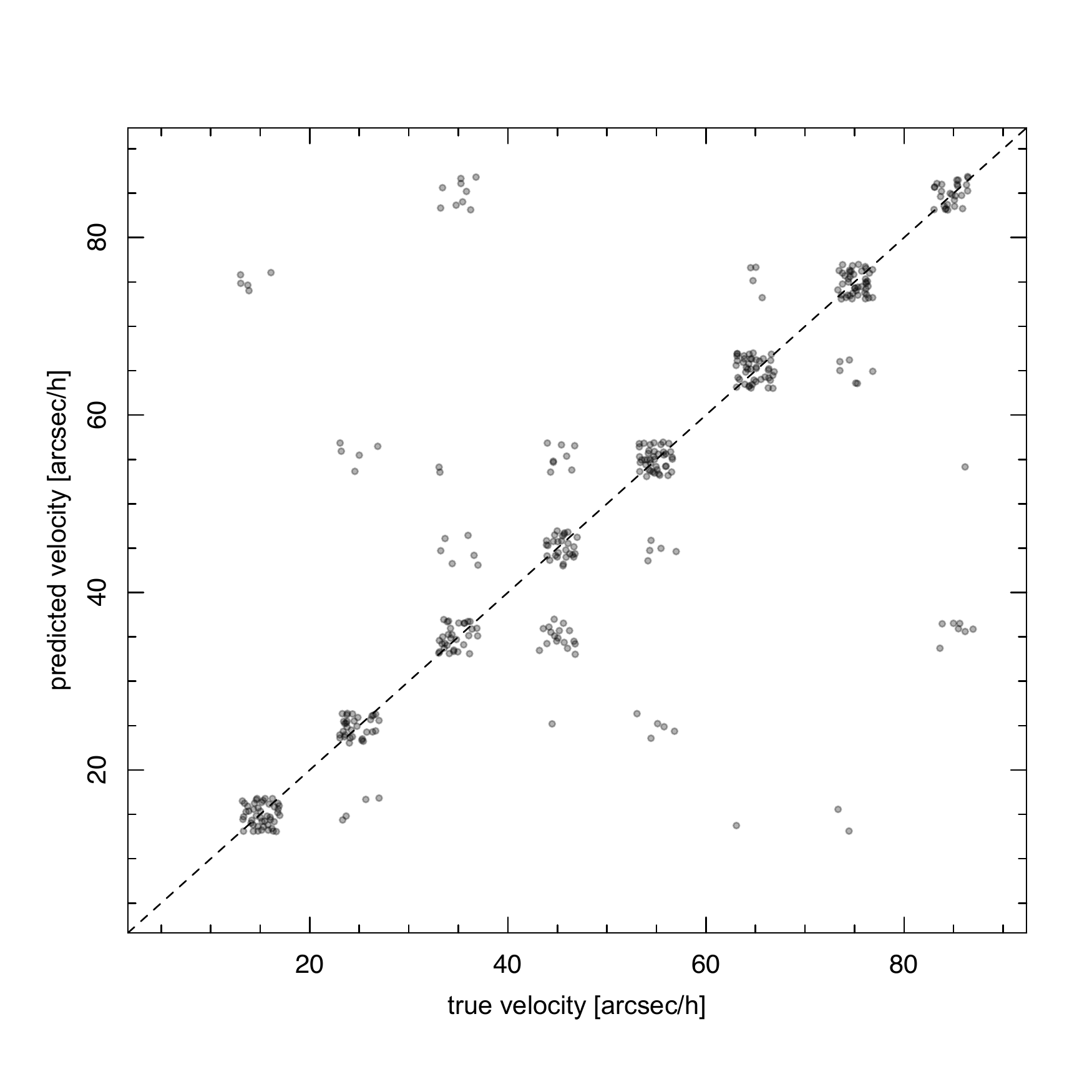}
\caption{\textit{Left:} Fraction of correctly predicted asteroid velocities against true velocity. The error bars show the standard error. \textit{Right:} Predicted velocity bin vs true velocity bin of asteroids. The dashed line represents equality, and the points are jittered in both x and y for clarity.
\label{Fig: frac_vel_pred}}
\end{figure*}

Convolutional neural networks can also be used to predict continuous quantities such as the velocities of asteroids. This is important for followup observations of SSOs. We repurpose the MobileNet architecture to predict the velocity bin of asteroids in the single dither images. The result trained in 10518 seconds with a 77$\%$ accuracy score on the testing data. 
Asteroids in the in 10-20 arcsec/h bin had the highest fraction of correct predictions (90.2$\%$) and whereas the those in the 40-50 arcsec/h scored the worst (only 51.9$\%$ were correct). It is an unlikely that the lowest velocity asteroids also had the largest missing fraction in classification due to bias, and more likely to be due to low number statistics of the test data ($\sim$ 50 SSOs per velocity bin). We found no trend for the accuracy of velocity prediction as a function of true velocity however, upon further investigation we found that the 40-50 arcsec/h asteroids that were incorrectly classified tended to be classified in the adjacent velocity bins (\autoref{Fig: frac_vel_pred}). Velocity prediction would be useful to further constrain the identification of asteroids as it would enable the prediction of an asteroid's location in consequent dithers, however this is beyond the scope of this paper.

\section{Conclusions}
\label{sec:conclusions}
We use 3 of the best deep convolutional neural nets currently available and apply transfer learning to 
retrain them for the classification asteroids. The neural nets are retrained on Euclid-like images, to identify 
asteroids and non-asteroids. The MobileNet model is found to be the best performing architecture 
on this dataset and is also the quickest CNN to train. Our model reaches top-accuracies of 94$\%$ 
and marginally increases to 96$\%$ when we include an additional 2 of Euclid's 4 dither images. The model is shown to further improve on the addition of augmentation (top accuracy 96$\%$ on the single dither images), on the other hand we find that using all 4 dithers or training scratch does not achieve high performance. We suspect that this is due to the limitations of the training dataset making it more difficult to constrain the model parameters.

 Our model is robust to the 
expansion of more labels. When the non-SSO class is replaced with classes; galaxies, stars and cosmic 
rays, the AUC (area under the receiver operating characteristic curve) score on asteroids decreases slightly by 2$\%$. This means that method has the potential to perform well on the classification of all objects in Euclid, provided that a more complex training data set of simulations were to be produced. This includes both astronomical objects and instrumental artefacts such as supernova, satellite trails and ghosts. Our research suggests that the models with and without dithering information are complementary to each other. We find that without using dithering information, our asteroid sample was more contaminated by stars, however including the dithering information the predicted asteroids were more contaminated by cosmic rays.

Our CNNs perform well on even the faintest asteroids but are more susceptible to missing the slowest moving asteroids. The true abundance of asteroids will be low, therefore to achieve a high purity and completeness sample we could preprocess the data with a method such as SExtractor \citep{Bertin1996} for an initial removal of stars, galaxies and cosmic rays. Nonetheless, the Euclid pipeline will likely remove the majority of cosmic rays. 

Finally, we show that the same technique can be applied to predict the velocities of the asteroids to 77$\%$ accuracy and with no obvious signs of bias. This opens up a large number of possibilities in the future for analysing astronomical data from Euclid and other big data volume missions such as LSST.

\section*{Acknowledgements}
The authors would like to thank Justin Alsing, David Hogg, Will Farr, Stephen Feeny, Michelle Lochner and Matej Kosiba for useful discussions. ML acknowledges a Postgraduate Studentship from the Science and Technology Facilities Council and an ESA Research Fellowship at the European Space Astronomy Centre (ESAC) in Madrid, Spain. We also thank the anonymous referee for the constructive comments.

\bibliographystyle{mnras}
\bibliography{references}

\appendix


\bsp	
\label{lastpage}
\end{document}